\documentclass{aa}
\usepackage{graphicx}
\usepackage{natbib}
\usepackage{amssymb,amsmath}
\bibpunct{(}{)}{;}{a}{}{,}
\usepackage[varg]{txfonts}
\usepackage{subfigure}
\usepackage{epstopdf}
\usepackage{float}
\usepackage{longtable}
\usepackage{hyperref}
\usepackage{color}
\usepackage{orcidlink}
\begin{document}

   \title{ Revealing Event Rate of Repeating Fast Radio Bursts }

   \subtitle{}

\author{Q. Pan
	\inst{ 1,7 }
	X. Y. Du
	\inst{1,7}
	\and
	Z. B. Zhang 
	\inst{ 1,* } 
	\and
	Y. F. Huang 
	\inst{2,3,*} 
	\and
	L. B. Li
	\inst{4,5,6,*} 
	\and
	G. A. Li
	\inst{1}
}

\institute{College of Physics and Engineering, Qufu Normal University, Qufu 273165, P. R. China.            \email{z\_b\_zhang@sina.com}  \label{inst1}
	\and
	School of Astronomy and Space Science, Nanjing University, Nanjing 210023, P. R. China.   \email{hyf@nju.edu.cn}\label{inst2}
	\and
	Key Laboratory of Modern Astronomy and Astrophysics (Nanjing University), Ministry of Education, P. R. China \label{inst3}
	\and
	School of Mathematics and Physics, Hebei University of Engineering, Handan 056005, P. R. China.   \email{lilongbiao@hebeu.edu.cn.}\label{inst4}
	\and
	Hebei Computational Optical Imaging and Photoelectric Detection Technology Innovation Center, Hebei University of Engineering, Handan 056005, P. R. China\label{inst5}
	\and
	Hebei International Joint Research Center for Computational Optical Imaging and Intelligent Sensing, Hebei University of Engineering, Handan 056005, P. R. China \label{inst6}
	\and
	These authors have contributed equally to this work. \label{inst7}	
}

   \date{Received; accepted:accepted for publication in $A\&A$}

\abstract{How the event rate of fast radio bursts (FRBs) evolves with redshift is a hot topic to explore their cosmological origin and the circum-burst environment. Particularly, it is urgent to know what the difference of event rates between repeating and non-repeating FRBs is. For the first time, we calculate the event rates of repeating FRBs detected by diverse telescopes at frequencies higher/lower than 1 GHz in this work. Luminosity and redshift are found to be positively correlated with a power law form for both high- and low-frequency FRBs, showing an obvious evolution of luminosity with redshift. Furthermore, we compare the differential luminosity and local event rate distributions of high- and low-luminosity FRBs at different frequencies. It is found that the event rates of these sub-samples of repeating FRBs similarly exceed the star formation rate at lower redshift than 1. Interestingly, we confirm with bootstrap method that the event rates of low-frequency FRBs exhibit different evolution patterns and are higher than that of high-frequency ones.}

   \authorrunning{ Pan et al. }            %author_head in even pages
   \titlerunning{ Statistical properties of repeating FRBs }  % title_head in odd pages
   \maketitle
https://ui.adsabs.harvard.edu/abs/2023MNRAS.524.1096L/abstract
%________________________________________________ sections below
%
\section{Introduction} \label{sec:intro}         %% first-level sections will be auto-capitalized

Fast radio bursts (FRBs) are mysterious bright transients at radio frequencies detected by ground-based radio telescope, with millisecond duration and extremely brightness temperature of $\sim$ 10$^{37}$ K. FRBs were first discovered in archival data of Parkes telescope \citep{Lorimer+07} and confirmed to be of an astrophysical origin by \cite{Thornton+13}. Since the report of the first repeating FRB 20121102A \citep[e.g.,][]{Spitler+16,Scholz+16} and the further identification of its host galaxy \citep[e.g.,][]{Chatterjee-17,Marcote-17,Tendulkar-17}, the current state of the FRBs' field is very active. To date, over 800 FRBs have been discovered, of which the majority have been detected only one time, and called as non-repeating FRBs or one-offs. Other sources are interestingly found to be repeating FRBs, of which the number has reached up to 67 \citep{Xu-23}, e.g., more than 2370 repeating bursts from FRB 20121102A have been reported \citep{LiDi+21}. Due to their cosmological origin, high dispersion measure (DM) and energetic nature, FRBs have acted as a probe of the cosmic web, ggalactic halos, baryons, etc \citep[e.g.,][]{RaviV-16,Prochaska-19,Macquart-20}.

It is difficult to clarify the progenitors and radiation mechanisms. Until now, only $\sim$ 10\% FRBs are conclusively associated with host galaxies. Dozens of theoretical models have been proposed (see \cite{Platts-19} for a review). Observationally, the actively star-forming host galaxy \citep{Tendulkar-17} and an extremely large Faraday rotation measure (RM) of FRB 20121102A \citep{Michilli-18}, and the detection of the Galactic FRB 20200428 \citep[e.g.,][]{CHIME/FRB-20, Bochenek-20, LiCK-21}, indicate that young magnetars might be sources of active repeating FRBs. Such a simple picture where all FRBs originate from young magnetars cannot explain the diverse emission and host galaxy properties displayed by most sources, including both repeaters and non-repeaters. Some other models, e.g., collisions between episodic magnetic blobs \citep{LiLB-18}, the orbital motion of a binary system \citep{LiQC-21}, orbit-induced spin precession, may also work.

It is worth noting that the luminosity function, burst rate and burst environment of FRBs can place important constraints on the physical origins and progenitor models of FRBs \citep[e.g.,][]{LuoR-18,LuoR-20,ZhangW-19,Bhattacharya-20,ZhangK-22}. \cite{Rane+16} have argued that the all-sky event rate of FRBs source could reach 4.4 $\times$ 10$^{3}$ d$^{-1}$ sky$^{-1}$ with fluence exceeding 4.0 Jy ms at 1.4 GHz, which is consistent with the theoretical event rate estimated with some progenitor models, such as the collisions of asteroids with neutron stars \citep{GengJ-15}. The event rates of FRBs located at high Galactic latitudes and low/mid-latitudes were considered to be different \citep{Vander+16}. Compared to non-repeating burst, repeating FRBs' burst rate, according to the observation of the Chinese Five-hundred-meter Aperture Spherical radio Telescope (FAST), could be as high as 0.1 to $\sim$100 hr$^{-1}$ \citep[e.g.,][]{LiDi+21,ZhangYK-22}. Generally, the host galaxies of FRBs are suggested to be accompanied by at least moderate rate of star formation, such as FRB 20121102A's host galaxy \citep{Tendulkar-17}. \cite{Yamanaka+24} found that the star formation rate (SFR) of FRB 20191001A host galaxy is different from that of other FRB host galaxies. The high SFR near FRB 20201124A indicates that the corresponding progenitor should be a newborn magnetar produced during the supernova explosion of a massive star progenitor \citep{Piro+21}. By using a CHIME 435 FRB sample, \cite{Zhang+23} proposed that the evolution of FRB population with redshift is consistent with, or faster than the SFR, which supports the hypothesis of FRBs originating from young magnetars. In addition, \cite{ZhangW-19} argued that FRBs may occur in low-metallicity environment, which is similar to long gamma-ray bursts (GRBs) and hydrogen-poor superluminous supernovae (SLSNe-I). Thus, FRBs are considered to originate from multiple progenitors across a diverse range of galaxy environments.

In this work, we collect the observational data of 65 repeating bursts from the Blinkverse Database \footnote{ Blinkverse: \url{https://blinkverse.alkaidos.cn/\#/overview} }, categorizing them into different samples based on the observational frequency bands and estimated luminosities, to study the luminosity distribution and the event rates of FRBs. The structure of this paper is organized as follows. In Section~\ref{sec:samples and method}, we introduce the repeating FRB sample and the theoretical method. Our results are presented in Section~\ref{sec:results}. Relevant discussion and conclusions are made in Section~\ref{sec:discussion and conclusions}. Here, a $\Lambda$CDM model with $H_{0}$ = 69.6 km $\cdot$ s$^{-1} \cdot$ Mpc$^{-1}$, $\Omega_M =0.286$ and $\Omega_\Lambda =0.714$ is adopted \citep{Bennett+14}.

%%%%%%%%%%%%%%%%%%%%%%%%%%%%%%%%%%%%%%%%%%%%%%%%%%

\section{sample and method}
\label{sec:samples and method}
\subsection{Samples}
\label{sec:samples}

In our work, we focus on the luminosity distribution and the event rate of repeating FRBs with respect to redshift $z$. As of December 2023, 65 repeating FRBs have been detected, which comprise our sample. We extract the observational data of these 65 repeating FRBs from Blinkverse Database. For each repeating FRB, all the repeating pulses detected by different ground-base radio telescopes are classified into high frequency (HF) and low frequency (LF) samples with a boundary of the observational central frequency at 1 GHz. As shown in Table \ref{table1}, Columns 1-3 present the name of repeating events, the number of each FRB source's repeating pulses detected, and DM values, respectively. By adopting the YMW16 model \citep{Yao+17}, we estimate the redshift $z$ of each repeater, as listed in Column 4 of Table \ref{table1}. According to $z$ and the luminosity distance $D_{L}(z) = \frac{c}{H_{0}} (1+z) \int_{0}^{z} \frac{dz}{\sqrt{\Omega_M (1+z)^{3} + \Omega_{\Lambda}}}$ \citep{Wright+06}, where the light speed $c$ is 3 $\times$ 10$^{8}$ m/s, we can adopt the method of \cite{LiXJ+21} to calculate the bolometric luminosity of each burst, i.e.

\begin{equation}
	L = 4\pi D^{2}_{L} S_{\nu} \nu_{c},
\end{equation}
where S$_{\nu}$ is the radio flux density at a central frequency of $\nu_{c}$. 

\begin{figure}[!h]
	\centering
	\includegraphics[width=\columnwidth]{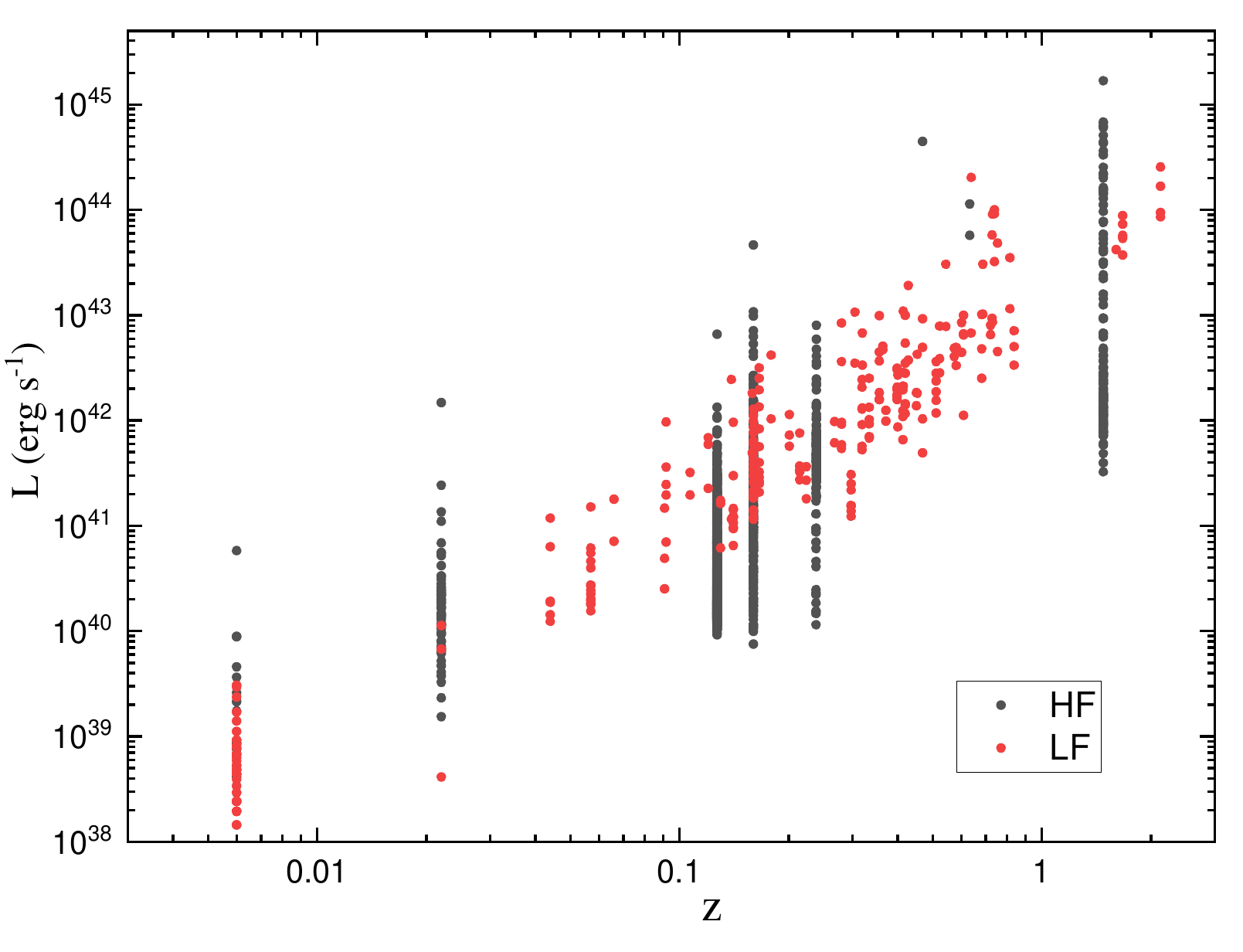}
	\caption{The relationship of luminosity $L$ vs. redshift $z$. The black dots and red dots refer to the repeating FRBs observed at central frequency $\textgreater 1$ GHz and $\textless 1$ GHz, i.e. HF and LF samples, respectively.}
	\label{fig.1}
\end{figure}

Fig. \ref{fig.1} illustrates the correlation between the luminositys and redshifts of the repeating FRBs. It can be seen that more than 50 \% of repeating FRBs are localized in redshift region of $z$ $\textgreater$ 0.1, and the number of HF category is obviously less than that of LF category. We also find that the redshift and the luminosity are positively correlated, which means the observed luminosities of the repeating FRBs evolve with redshift in evidence. Moreover, the HF and LF are homegenously  distributed on the whole sky.

For HF and LF classes, we extract the detected bursts with the maximum and minimum peak flux densities, corresponding to the highest luminosity (HL) and lowest luminosity (LL). In this way, four repeating sub-samples, namely HF-HL, HF-LL, LF-HL, LF-LL, are built and denoted in Column 5 of Table \ref{table1}. The central frequency $\nu_c$, flux density $S_\nu$, pulse width, fluence $F_{\nu}$, and bolometric luminosity are respectively listed in Columns 6-10. The main instrument parameters of telescopes are listed in Table \ref{table2}.

\subsection{Method}
\label{sec:event rate}

Following \cite{Sun+15}, we assume that the events with the same luminosity share similar other properties (e.g., spectral properties and detector parameters) and ignore possible redshift evolutions of the luminosity functions of our sub-samples. We define the local specific event rate density (local event rate density per unit luminosity) at a specific luminosity as

\begin{equation}
	\label{eq.2}
	\rho_{0,L} = \frac{4 \pi}{\Omega T} \frac{1}{\rm{ln} 10} \frac{1}{g(L)} \frac{1}{L} \frac{\Delta N}{\Delta \rm{log} L},
\end{equation}
where $\Omega$ and $T$ are respectively the field of view and operation time for a telescope as shown in Table \ref{table2}, $\Delta N$ is the number of events detected in a finite logarithmic luminosity bin of $\Delta$log$L$, and the factor $g(L)$ is determined by

\begin{equation}
	\label{eq.3}
	g(L) = \int_{0}^{z_{max}} \frac{f(z)}{1+z} \frac{dV(z)}{dz} dz,
\end{equation}
in which the redshift-dependent specific comoving volume can be written as

\begin{equation}
	\frac{dV(z)}{dz} = \frac{c}{H_0} \frac{4\pi D^{2}_{L}}{(1+z)^{2} \sqrt{\Omega_M (1+z)^{3} + \Omega _{\Lambda}}} ,
\end{equation}
and $f(z)$ is the redshift-dependent term, whose form depends on the properites of the transient type. Here, we assume that the distributional function $f(z)$ of FRBs has the same form as the undelyed SFR, i.e.

\begin{equation}
	f(z) = \left[ (1+z)^{a\eta} + \left( \frac{1+z}{B} \right)^{b\eta} + \left( \frac{1+z}{C} \right)^{c\eta}   \right] ^{1/\eta},          
\end{equation}
with the best fitting parameters of $\eta=-10$, a=3.4, b=-0.3, c=-3.5, B$\approx$5000 and C$\approx$9 \citep{2008ApJ...683L...5Y}. With the flux sensitivity of $F_{\rm th}$, the maximum redshift $z$$_{max}$ in Eq. (\ref{eq.3}) for an event with a given luminosity $L$ is constrained via $L= 4 \pi D^{2}_{L} (z_{\rm max})F_{\rm th}$. Meanwhile, the local event rate density above a given luminosity $L$ can be extimated with 
\begin{equation}
	\label{eq.7}
	\rho_{0,>L} = \sum_{{\rm log} L}^{ {\rm log} L_{\rm max}} \frac{4\pi}{\Omega T} \frac{1}{{\rm ln} 10} \frac{1}{g(L)} \frac{\Delta N}{\Delta {\rm log} L} \frac{\Delta L}{L} = \sum_{{\rm log} L}^{{\rm log} L_{\rm max}} \rho_{0,L} \Delta L ,
\end{equation}
where $L_{\rm max}$ denotes the maximum luminosity of FRBs in our samples. Taking the minimal luminosity $L$$_{min}$, one can derive the reshift-dependent event rate of FRBs as

\begin{equation}
	\label{eq.8}
	\rho(z) = \frac{4\pi}{\Omega T} \frac{dN}{dz} \left( \frac{dV}{dz} \right)^{-1} (1+z) \left( \int_{L_{min}}^{L_{max}} \Phi (L) dL \right) ^{-1} ,
\end{equation}
where $\Phi(L)$ is the luminosity distribution function that can be described by a single power-law (SPL) form of $\Phi(L) \propto L^{-\alpha_1}$ in some cases. Alternatively, it can be expressed by a smoothly broken power-law (BPL) function \citep{2019ApJS..245....1T} as
\begin{equation}
	\label{eq.9}
	\Phi(L) \propto \left[ \left( \frac{L}{L_{b}} \right)^{\omega \alpha_{1}}  +  \left( \frac{L}{L_{b}} \right)^{\omega \alpha_{2}}  \right] ^{-1/\omega},
\end{equation}
where $\alpha_{1}$ and $\alpha_{2}$ are respectively power-law indices before and after the broken luminosity $L_{b}$, and $\omega$ is the smoothness parameter featuring the sharpness of the $\gamma$-ray luminosity distribution functions.

%\twocolumn

%%%%%%%%%%%%%%%%%%%%%%%%%%%%%%%%%%%%%%%%%%%%%%%%%%%%%%%%%%

\section{results}
\label{sec:results}
\subsection{Luminosity and redshift distributions}

\begin{figure}[!h]
	\centering
	\includegraphics[width=\columnwidth]{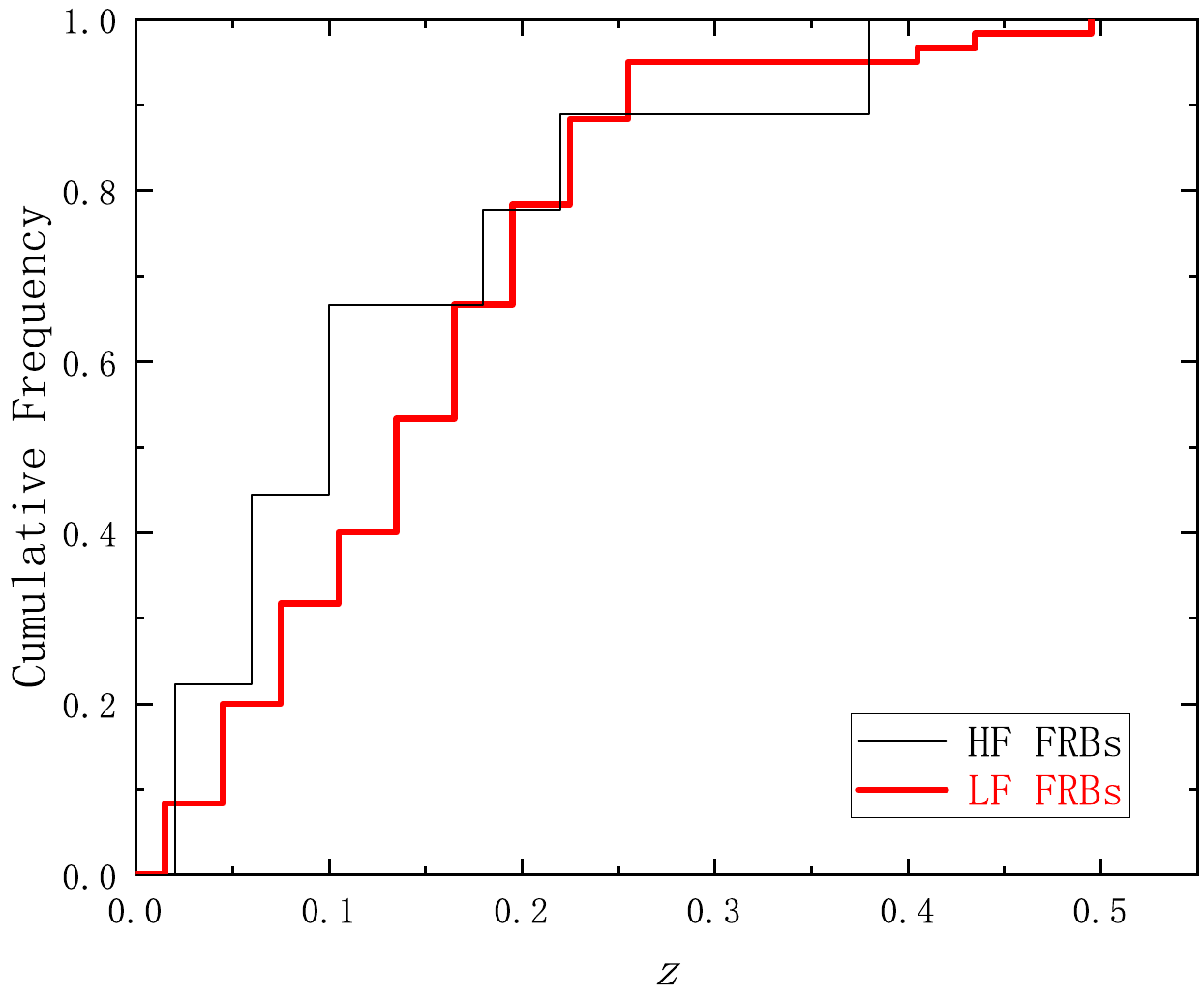}
	\includegraphics[width=\columnwidth]{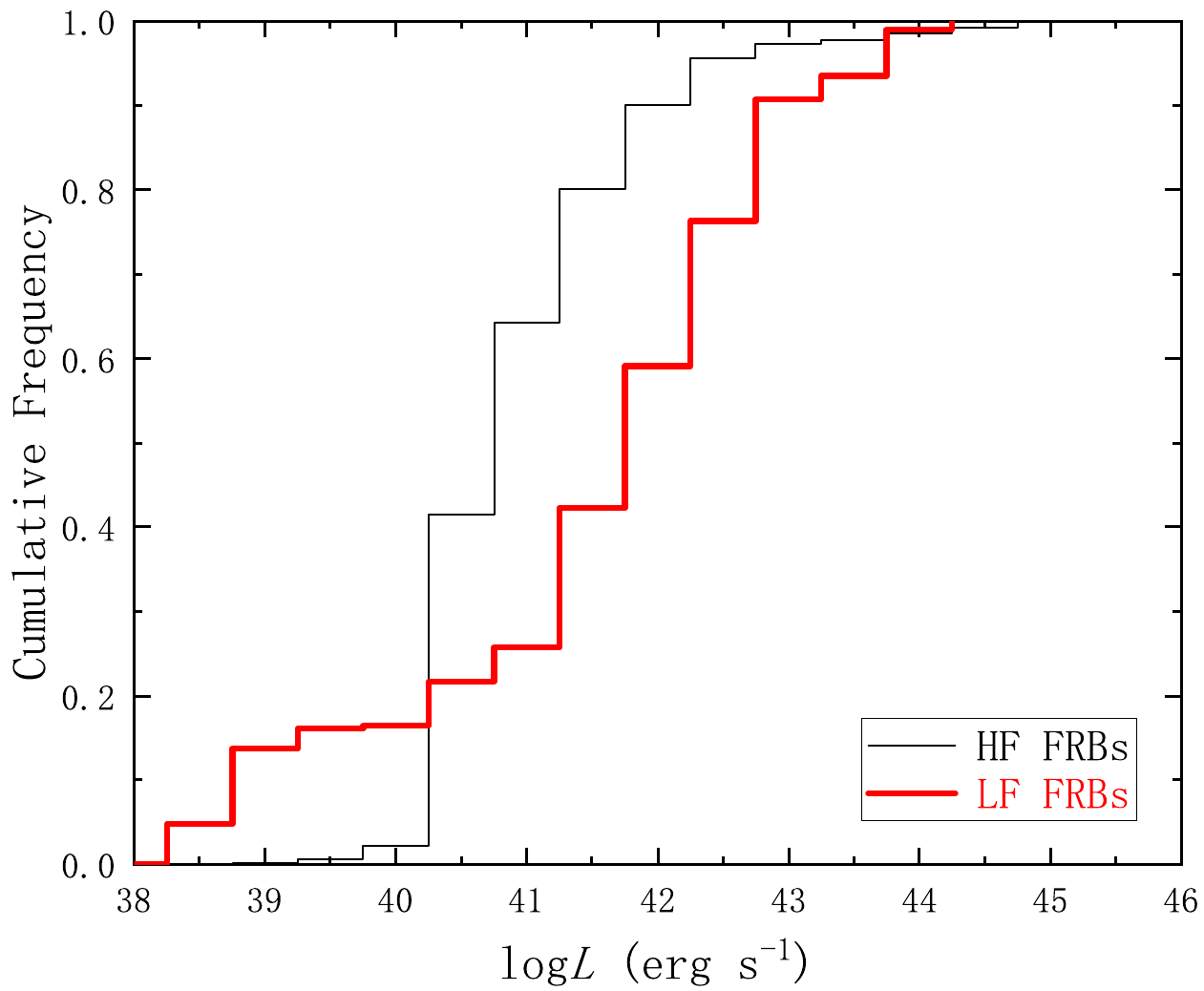}
	\caption{Cumulative redshift (top panel) and luminosity (bottom panel) distribitons of high-frequency (thin line) and low-frequency (thick line) repeating FRBs. 
		\label{fig.2}   }
\end{figure}

\begin{figure}[!h]
	\centering
	\includegraphics[width=\columnwidth]{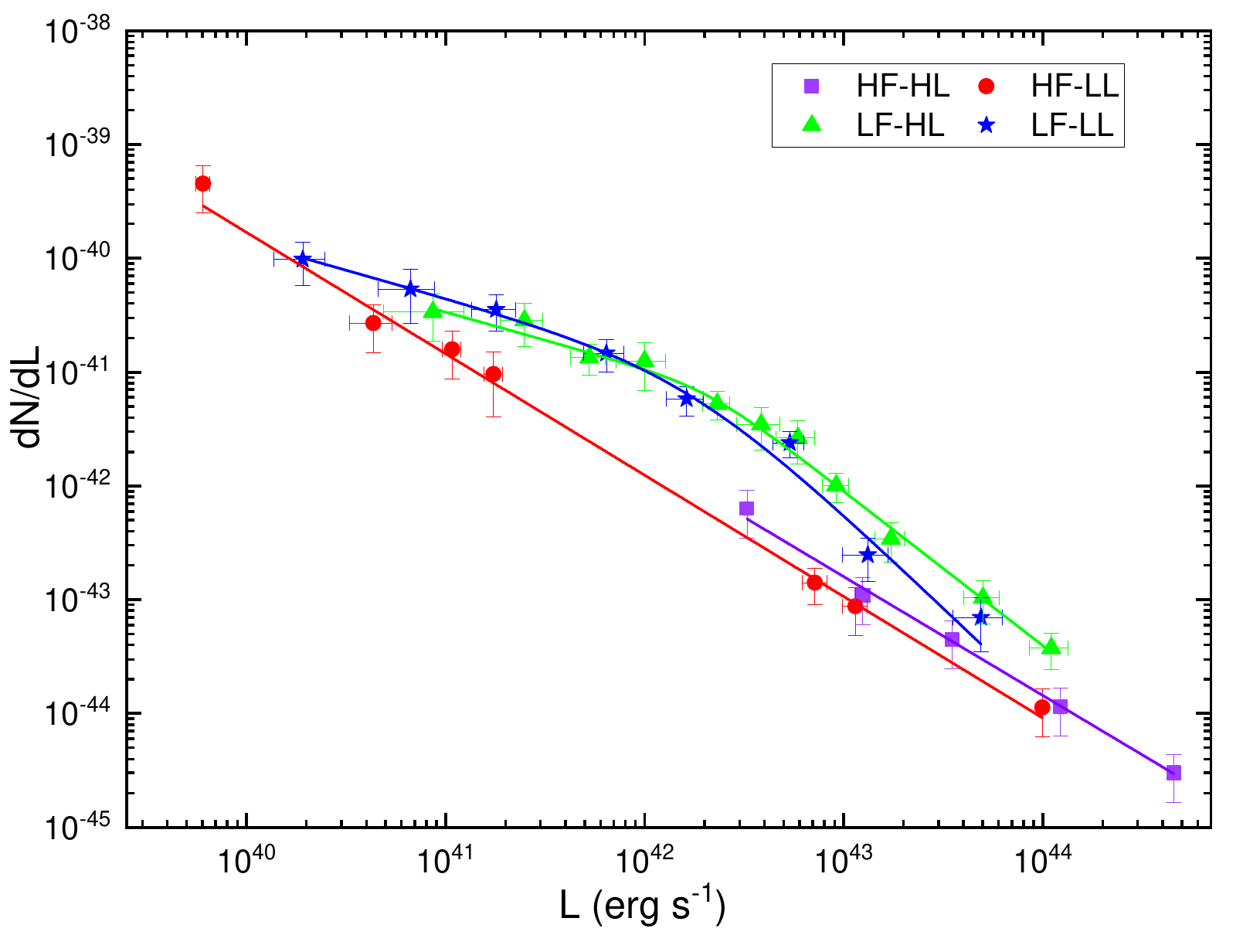}
	\caption{Differential luminosity distributions of HF-HL (purple square), HF-LL (red circle), LF-HL (green triangle) and LF-LL (blue star) repeating FRBs, respectively. The solid lines represent the best fits with either SPL or BPL models.
		\label{fig.3}   }
\end{figure}

Firstly, we display the distributions of luminosity and redshift at different frequecies in Fig. \ref{fig.2}, where a K-S test returns $D=0.35$ (less than the critical value of $D_{\alpha}=0.49$) for the redshift distributions and $D=0.41$ (smaller than $D_{\alpha}=0.09$) for the luminosity distributions at a significance level of $\alpha=0.05$, showing that the redshift distributions are identical while their luminosity distributions are diverse. Secondly, we compare the differentical luminosity distributions of four sub-samples of HF-HL, HF-LL, LF-HL and LF-LL FRBs in Fig. \ref{fig.3}, from which we notice that the luminosity distributions of the two HF FRB samples can be well fitted with the SPL model, while the luminosity distributions of the two LF FRB samples can be better fitted with the BPL model. The best fitting parameters are given in Table \ref{table3}, from which we conclude that the luminosities of HF and LF repeating FRBs are differently distributed. However, the mean luminosity of HF-HL FRBs are abount two orders of magnitude larger than that of HF-LL FRBs. On the contrary, the luminosities of LF-HL and LF-LL FRBs are quite comparable.

%%%%%%%%%%%%%%%%%%%%%%%%%%%%%%%%%%%%%%%%%%%%%

%%%%%%%%%%%%%%%%%%%%%%%%%%%%%%%%%%%%%%%%%%%%%

\subsection{Local event rates}

\begin{figure}[!h]
	\centering
	\includegraphics[width=\columnwidth]{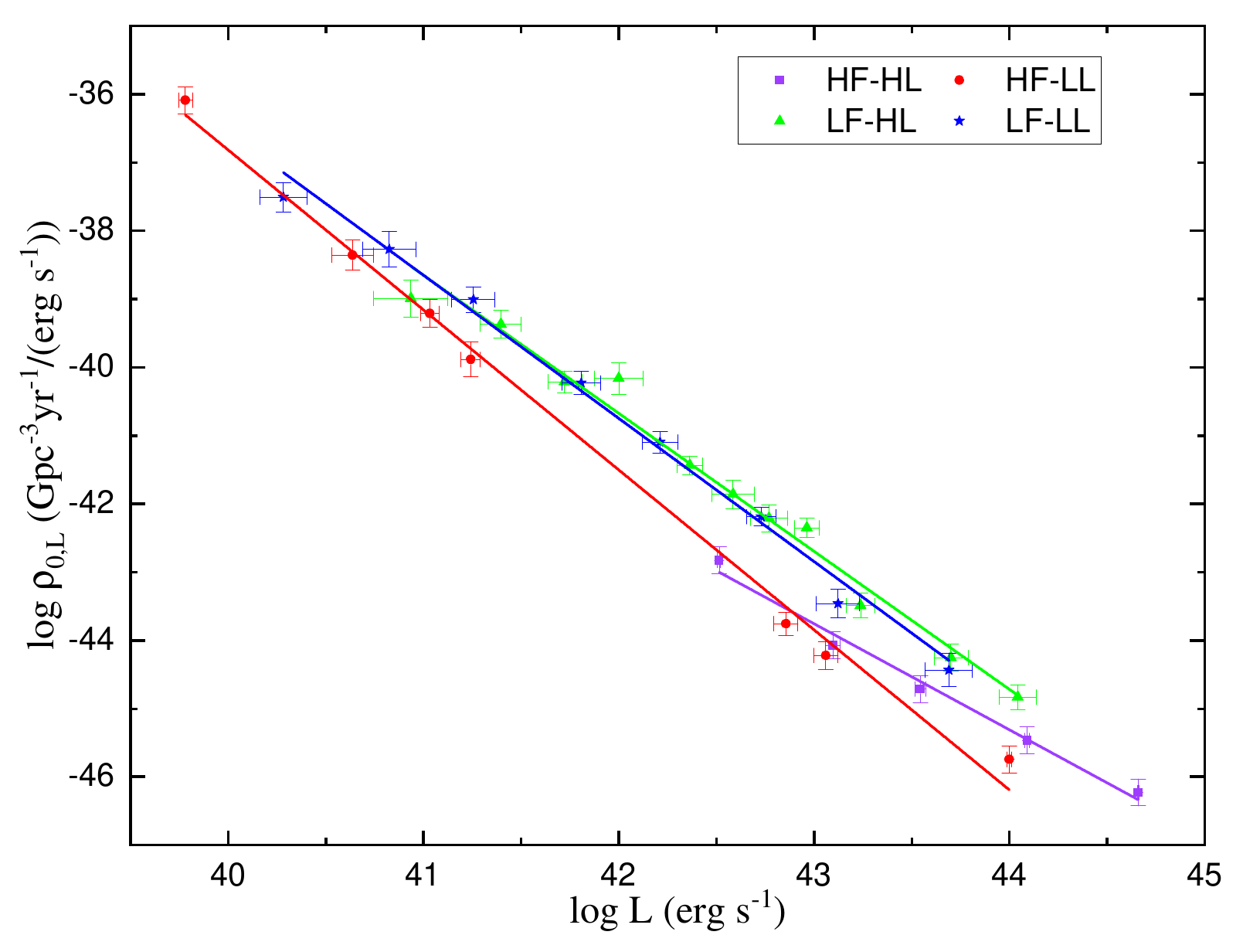}
	\includegraphics[width=\columnwidth]{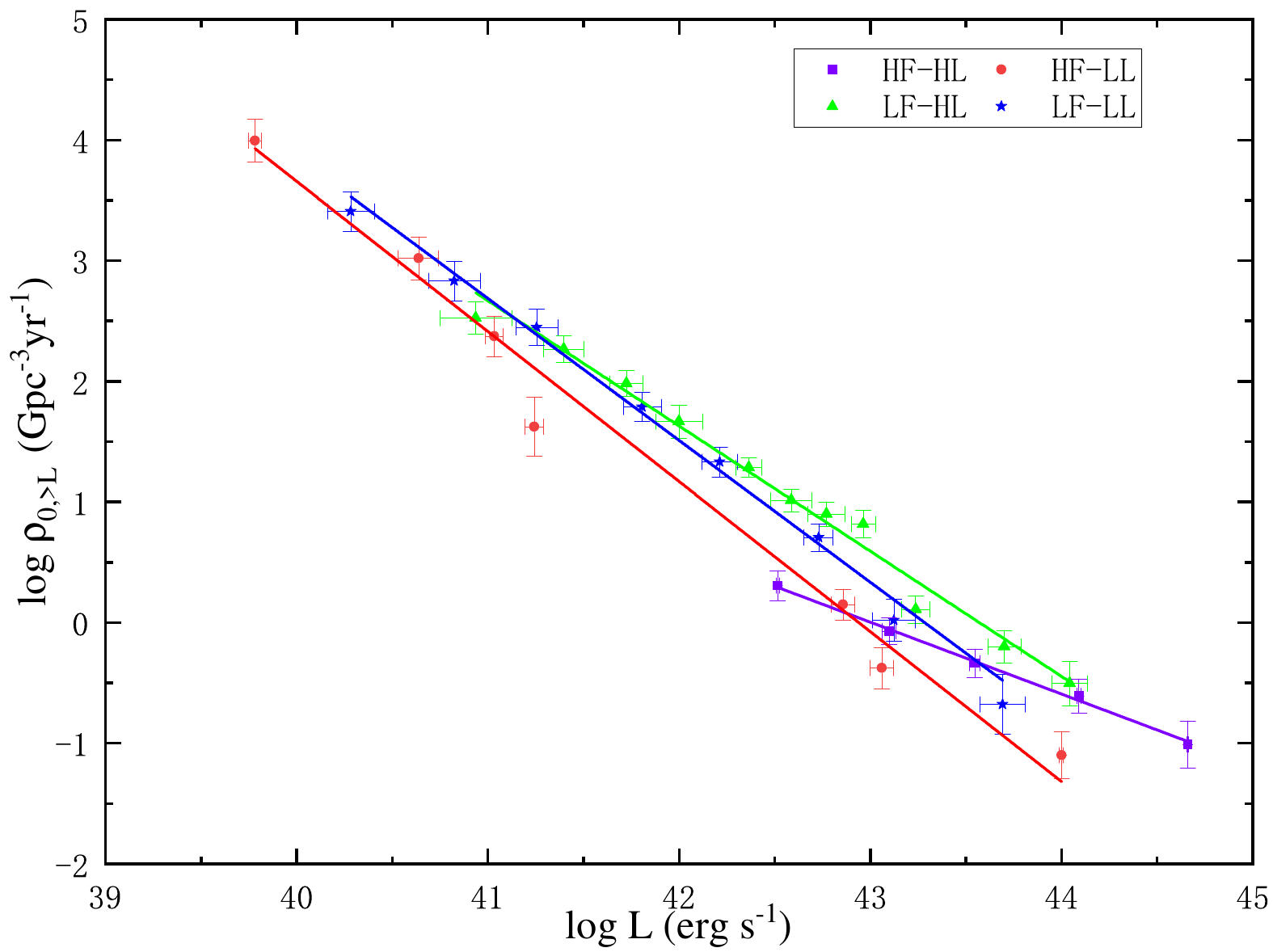}
	\caption{The relations of $\rho_{0,L}$ on top panel and $\rho_{0,>L}$ on bottom panel with $L$ for different types of repeating FRBs. The solid lines represent the best fits with the SPL form to each FRB sample.
		\label{fig.4}     }
\end{figure}

We now estimate the local event rate densities, $\rho_{0,L}$ and $\rho_{0,>L}$, of our four sub-samples and plot them against the luminosity in Fig. \ref{fig.4}. The relations between $\rho_{0,L}$ ($\rho_{0,>L}$) and $L$ are well fitted by the SPL function with power-law indices of $-1.55\pm0.09$ ($-0.59\pm0.02$), $-2.34\pm0.07$ ($-1.24\pm0.05$), $-2.02\pm0.08$ ($-1.03\pm0.04$) and $-2.09\pm0.08$ ($-1.17\pm0.04$) for HF-HL, HF-LL, LF-HL and LF-LL sub-samples, respectively. It is worth emphasizing that the power-law index of HF-HL FRBs is obviously different from those of other kinds of FRBs no matter which local event rate is considered.

To compare the $\rho_{0,L}$ or $\rho_{0,>L}$ distributions between different sub-samples, a two-dimensional Kolmogorov–Smirnov (2D K-S) test is performed. The relevant testing results are presented in Table \ref{table4}, in which the $\rho_{0,L}$ distributions of HF-LL, LF-HL and LF-LL FRB sub-samples are identical but significantly differ from that of HF-HL FRBs. Interestingly, the $\rho_{0,>L}$ distributions of different types of FRBs exhibit similar patterns. Note that the luminosity evolution patterns of both local event rates are also supported by the above power-law indices of these sub-samples. The discrepncy of the $\rho_{0,L}$ or $\rho_{0,>L}$ between  HF-HL and other sub-samples implies that the local event rates could be biased by not only the observational frequencies but also the luminosity values.

%%%%%%%%%%%%%%%%%%%%%%%%%%%%%%%%%%%%%%%%%%%%%
%%%%%%%%%%%%%%%%%%%%%%%%%%%%%%%%%%%%%%%%%%%%%

\subsection{The observed FRB rates versus the SFRs }

\begin{figure}[!h]
	\centering
	\includegraphics[width=\columnwidth]{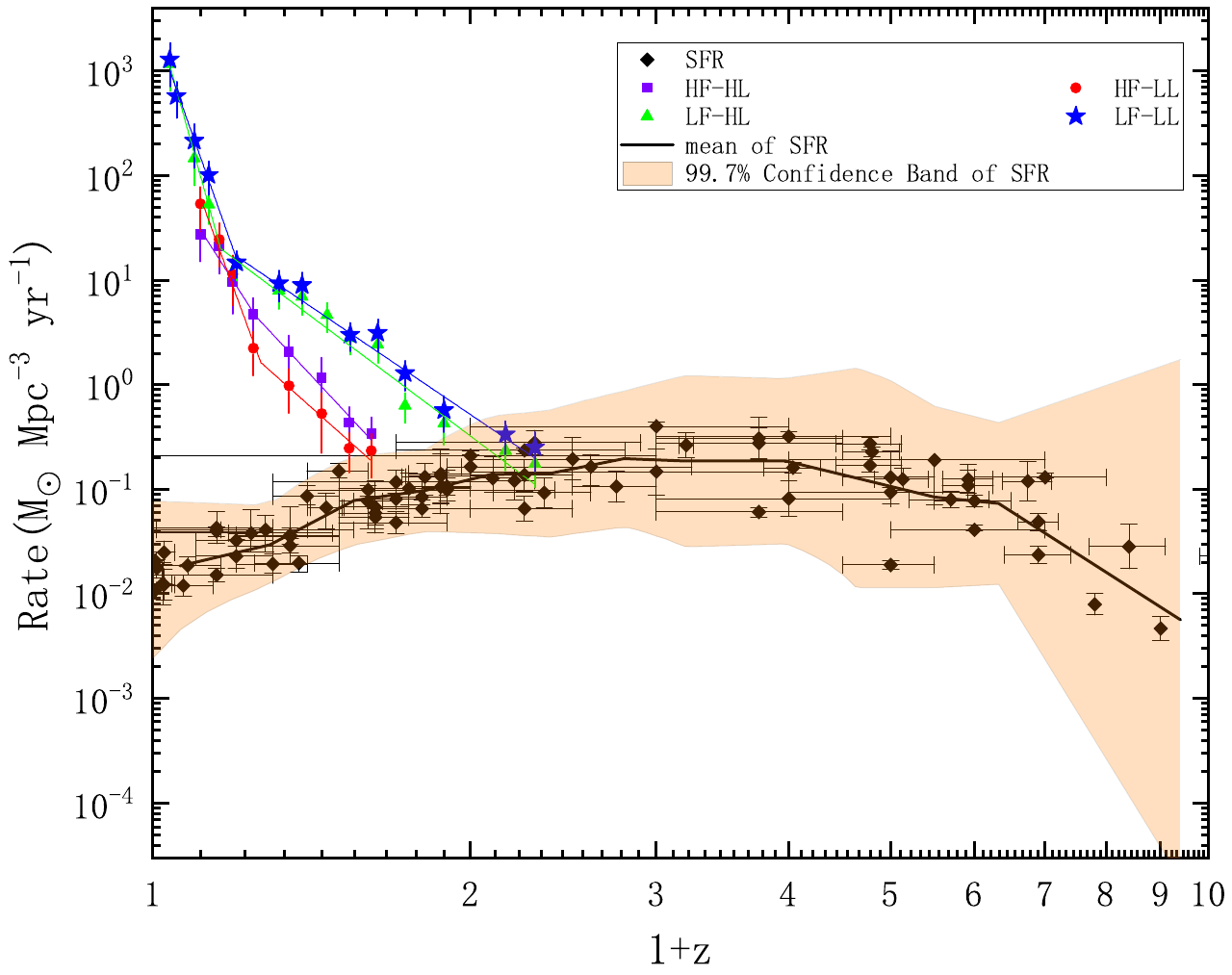}
	\caption{Comparison between four sub-samples of FRB rates and the SFR (filled diamonds). The black solid line and shade regions represent the means of the SFR data \citep{Hopkins+04} together with 3$\sigma$ confidence levels.	All symbols are illustrated in the insert.	\label{fig.5}     }
\end{figure}

Using Eq. (\ref{eq.8}) in Sec. \ref{sec:event rate}, we have calculated the redshift-dependent event rates of four FRB sub-samples, and compared these event rates with the SFR in Fig. \ref{fig.5}. It is found that the event rates of the four FRB sub-samples are significantly higher than the SFR at a lower redshift region of $z<1$, supporting the speculation of FRBs associated with older star populations \citep{Hashimoto+22,2025A&A...698A..18Z}. Furthermore, we notice that the event rates of high- and low-frequency FRBs evolve with redshift in a different way although all kinds of FRBs have smaller rates at higher redshift and they decease towards higher redshift with a broken power-law form as

\begin{equation}
	\label{eq.10}
	R(z) = \left\{
	\begin{array}{rcl}
		A((1+z)/B)^{C} &, & {(1+z) < B}, \\
		A((1+z)/B)^{D} &, & {(1+z) \geq B} ,
	\end{array} \right.
\end{equation}
where $A$, $B$, $C$ and $D$ are the free parameters. The best fitted parameters have been listed in Table \ref{table5}. It can be seen that the observed event rates of four FRB samples drop quickly at lower redshift and then decline at higher redshift slowly. In addition, high- and low-frequency FRB rates evolve with redshift in different modes in that the low-frequency FRB rates are relatively larger and slower than the high-frequency ones.

%%%%%%%%%%%%%%%%%%%%%%%%%%%%%%%%%%%%%%%%%%%%%%%%%

\begin{figure}[!h]
	\centering
	\includegraphics[width=\columnwidth]{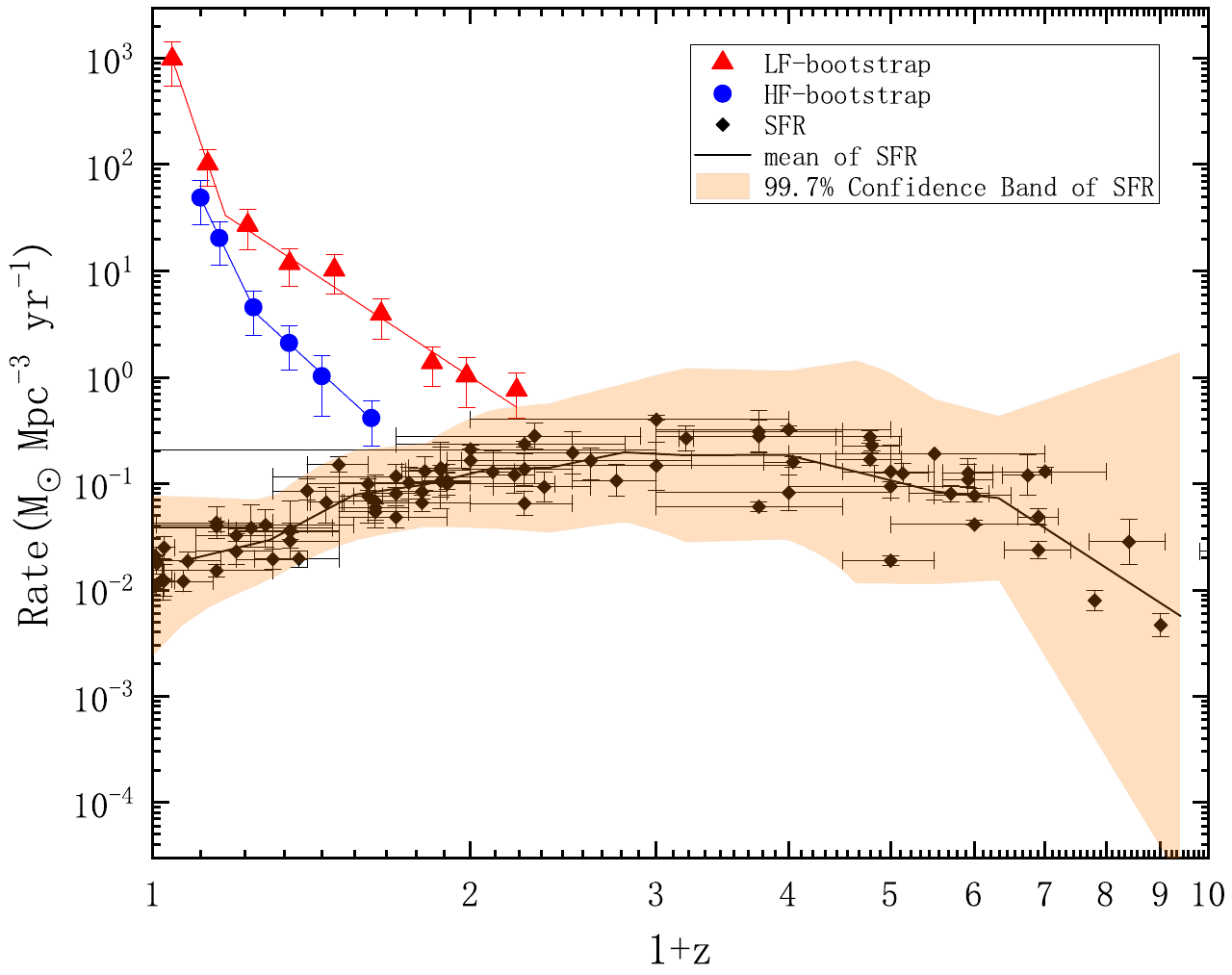}
	\caption{Comparison of the event rates of the Bootstrapped HF (filled circles) and LF (filled triangles) FRB samples with the SFR. All SFR symbols are the same as in Figure \ref{fig.5}. 
		\label{fig.6}      }
\end{figure}

\begin{figure}[!h]
	\centering
	\includegraphics[width=\columnwidth]{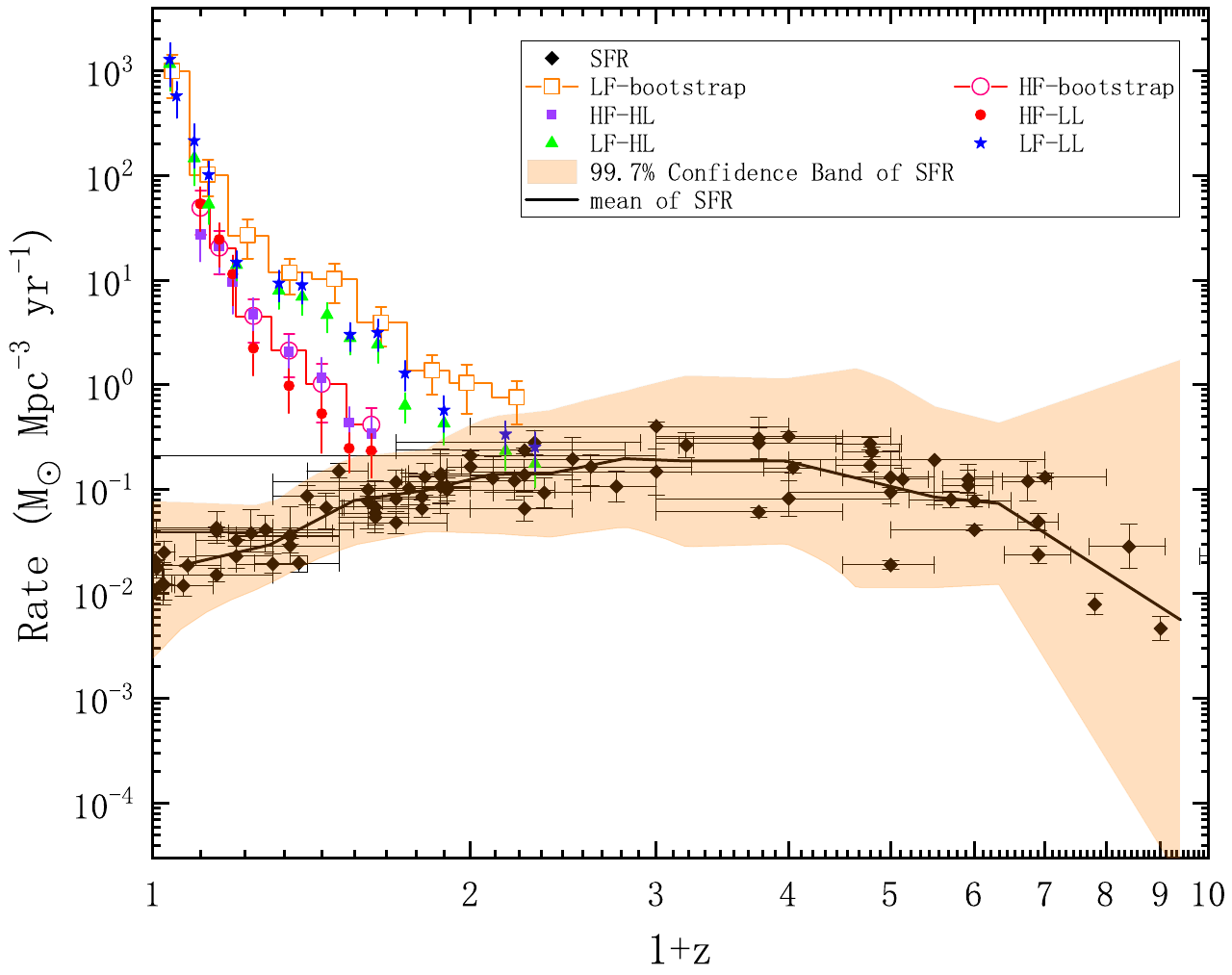}
	\caption{Comparison between the event rates of  four FRB sub-samples, two boostrapped samples (thick lines) and the SFR (fille diamonds). All symbols are the same as those in Figure \ref{fig.5}. 
		\label{fig.7}     }
\end{figure}

Considering the diversity or limitation of sub-burst numbers in repeating FRBs, we perform a bootstrap method to resample 1000 times of pulses from the original observations, ensuring the same number of sub-bursts for each repeating FRB. Similarly, we divide the resampled data into low- and high-frequency sub-samples with a boundary of $\nu_c$=1 GHz and then compute their bolometric luminosities and event rates as shown in Figs. \ref{fig.6} and \ref{fig.7}. It is evidently verified with the bootstrap strategy that the event rates of repeating FRBs do exceed the SFR at lower redshift region of $z<1$ and decrease with redshift variously for distinct frequencies, which roughly concides with the evolutionary trends of the four sub-samples of FRBs observed directly. The simulated FRB rates at different frequencies can also be fitted with Eq. (\ref{eq.10}) and parallelled with other observed FRB rates in Table \ref{table5}. Very interestingly, we find that all kinds of FRB rates change suddenly at a redshift of $z\approx0.2$, no matter which frequency is taken into account. This also demonstrates that the event rates of low- and high-frequency FRBs are comparable at lower redshift but become different at higher redshift. In contrast, the low-frequency FRB rates are larger than the high-frequency ones at farther distances.

%%%%%%%%%%%%%%%%%%%%%%%%%%%%%%%%%%%%%%%%%%%%%

\section{Conclusion and discussions}
\label{sec:discussion and conclusions}

In this work, we have systematically studied the redshift and luminosity distributions together with the event rate of repeating FRBs. In terms of multiple sub-bursts in each repeating source, we divided these sub-bursts into four classes, i.e., HF-HL, HF-LL, LF-HL and LF-LL sub-samples based on the central frequency and luminosity of these events. In particular, we focus on their  properties such as differential distribution of luminosity, and evolution of repeating FRB rates with both redshift and luminosity, and the connection between FRB rate and SFR. The following conclusions can be drawn:

\begin{itemize}
	\item The redshift distributions of HF and LF repeating FRBs are identical, while the luminosities of FRBs at different frequencies are diversely distributed. 
	\item The differential luminosity distributions of high-frequency (HF-HL and HF-LL) FRBs can be well fitted by the SPL function, while the  low-frequency (LF-HL and LF-LL) FRBs are better described by the BPL form. The power-law indices are independent of the luminosities.
	\item We have calculated the local event rates of the four types of repeating FRBs at/above a specific luminosity and found that they are related with the luminosity through a SPL form. The HF-HL FRBs have a comparatively lower local event rate slowly declining with the luminosity. 
	\item The derived event rates of the four sub-samples of repeating FRBs are found to evolve with redshift in a BPL form, decreasing with redshift sharply at lower redshift and decreasing slowly at higher redshift instead, which is further verified by the bootstrap resampling method.
	\item We find that all kinds of event rats of repeating FRBs exceed the SFR significantly at a redshift lower than 1. In comparison, the LF FRBs have relatively larger event rates at higher redshift range despite of their luminosity difference, which enables the LF FRBs to be more detectable. 
\end{itemize}

Based on the analysis of repeating FRBs with diverse luminosities measured at different frequencies, we can conclude that the most important characteristic on classifcation is  frequency rather than luminosity, even though the differential distribution of luminosities and local event rate densities show slight deviation between HF-HL and HF-LL FRB samples. This may demonstrate that the HF-HL sub-sample could constitute an isolated population, which resembles the HL long GRBs \citep[][Liu et al. 2025]{DongXF+23}, the Blazars (Rong et al. 2025) and the one-off FRBs \citep{2025A&A...698A..18Z}. Our findings are somethat consistent with \cite{2025A&A...698A..18Z} who had analyzed the event rate of non-repeating FRBs with non-parametric method and found that the HL FRBs follow the SFR and the LL FRBs deviate contrarily. Note that \cite{2025A&A...698A..18Z} had utilized not the flux-limited but the volume-limited FRB samples. However, we haven't obtained the trend of HL FRB rate matching the SFR. This may be attributed to the fact that our samples were also biased by the instrumental effect and/or the sample contamination more or less. Regardless of the differences between repeating and ``one-off'' FRBs, their observational properties are very similar or seriously overlapped \citep[e.g.][]{LiXJ+21,ZhangK-22,2025A&A...698A..18Z}, implying that all FRBs could be repeating in nature.

The origin of low-redshift excess of some burst rates with respect to the SFR is still controversial. \cite{DongXF+23} argued that the discrepancy is caused by the diversity of luminosities, in other words, the event rates of low-luminosity long GRBs exceed the SFR while the HL LGRBs match the SFR history excellently. \cite{2024ApJ...963L..12P} found that some LGRBs with higher event rate have a tendency aligning with a delayed SFR, which indicated that the low-redshift LGRBs may originate from a compact binary merger. It is noticeable that the LL LGRBs may have larger redshifts and those HL LGRBs might have smaller redshifts although most LL LGRBs have lower redshifts, similar to the majority of SGRBs. In practice, \cite{2025arXiv250601013D} proved by analyzing multi-satellite data that the SGRB rate always decreases with redshift and exceeds the SFR for not only the delayed but also the undelayed SFR models. \cite{Hashimoto+22} argued that if an older population or black holes produce FRBs, the event rate of FRBs will decrease towards higher redshifts. Low-luminosity FRBs may originate from binary star merger processes \citep{Hashimoto+22}, with smaller stellar masses and older stellar populations, which is consistent with our results presented in \cite{2025A&A...698A..18Z}. In short, the important findings in this work indicate that  repeating FRBs can be categorized into sub-groups by considering whether the central frequency is above 1 GHz or not and the two sub-groups have different luminosity distributions, event rate densities as well as physical origins, which will be hopefully testified by more observations of multi-messengers at multi-wavelengths.

%%%%%%%%%%%%%%%%%%%%%%%%%%%%%%%%%%%%%%%%%%%%%%%%%%%%%%%%%%%%%%

\begin{acknowledgements}
This work is supported by the National Natural Science Foundation of China (grant Nos. U2031118, 12041306 and 12233002), by the National SKA Program of China grant Nos. 2020SKA0120300, by the National Key R\&D Program of China (2021YFA0718500). Y.F.H. also acknowledges the support from the Xinjiang Tianchi Program.
\end{acknowledgements}
%%%%%%%%%%%%%%%%%%%%%%%%%%%%%%%%%%reference%%%%%%%%%%%%%%%%%%%%%%%%%%%%%%%%%%%%%%%%%%%%%%%%%%%%%%%%%%%%%%%%%%%%%%%%%%
\bibliographystyle{aa}
\bibliography{reference}
%%%%%%%%%%%%%%%%%%%%%%%%%%%%%%%%%reference 

%%%%%%%%%%%%%%%%%%%%%%%%%%%%%%%%%%%%%%table1%%%%%%%%%%%%%%%

\onecolumn

\begin{longtable}{cccccccccc}  
	\caption{ Observed and derived parameters of Repeating FRBs \label{table1}  }\\  
	\hline 
	{FRBs} & {N} & {DM} & {$z$} & {sub-samples} & {$\nu_{c}$} & {S$_{\nu}$} & {Width} & {F$_{\nu}$} & {$L$} \\
	{} & {} & {(pc cm$^{-3}$)} & {} & {} & {(GHz)} & {(Jy)} & {(ms)} & (Jy ms) & {( 10$^{41}$erg s$^{-1}$ )}  \\
	\hline
	\endfirsthead  % first page head
	\caption{ Observed and derived parameters of Repeating FRBs (continued) }\\  % Continued Table
	\hline
	{FRBs} & {N} & {DM} & {$z$} & {sub-samples} & {$\nu_{c}$} & {S$_{\nu}$} & {Width} & {F$_{\nu}$} & {$L$} \\
	{} & {} & {(pc cm$^{-3}$)} & {} & {} & {(GHz)} & {(Jy)} & {(ms)} & (Jy ms) & {( 10$^{41}$erg s$^{-1}$ )}  \\
	\hline
	\endhead  % continued page head
	\hline \multicolumn{10}{r}{ (continued) } \\  % page foot
	\endfoot  % each page foot
	\hline  
	\endlastfoot % last page foot
	
	20121102A 	& 2370  & 557.4 & 0.239 & HF-HL  & 6 &  $0.76\pm0.01$  & 1.43  & 0.61$\pm$0.02  & 79.89 \\
	&   &    &   &  HF-LL  & 6 &  $0.018\pm0.005$ & 2.15  & 0.026$\pm$0.013 & 1.87 \\
	20171019A 	&  7  & 460.8 & 0.469 & HF-HL  & 1.3 &  40.5  & 5.4$\pm$0.3  & 219$\pm$5  & 4417.65 \\
	&   &    &   &  HF-LL  & 1.3 &  40.5 & 5.4$\pm$0.3  & 219$\pm$5 & 4417.65 \\
	&   &    &   & LF-HL  & 0.6 & 1.82$\pm$0.98 & 7.6$\pm$1.6  & 22.6$\pm$8.4 & 91.84 \\
	&   &   &   &  LF-LL  & 0.82 & $\ldots$ & 5.2$\pm$0.8 & 0.37$\pm$0.05 & 4.91 \\
	20180301A	& 69  & 522 & 0.238 & HF-HL  & 1.25 &  0.105  & 4.1$\pm$0.06  & $\ldots$  & 2.27 \\
	&   &  &   &  HF-LL  & 1.25 &  0.005 & 5.5$\pm$0.6  & $\ldots$ & 0.11 \\
	20180916B	& 290  & 349 & 0.006$^{a}$ & HF-HL  & 1.7 & $\ldots$ & 0.06  & 2.53  & 0.58 \\
	&  &   &   &  HF-LL  & 1.7 & $\ldots$ & 2.4  & 0.72 & 0.0041 \\
	&  &  &  &  LF-HL  & 0.6 & 6.3$\pm$1.9 & 1$\pm$0.05  & 37$\pm$9 & 0.031 \\
	&  &  &  &  LF-LL  & 0.6 & 0.3$\pm$0.2 & 4.2$\pm$0.4 & 1.3$\pm$0.3 & 0.0015 \\
	20190520B	& 121  & 1204.7  & 1.478 & HF-HL  & 6 & $\ldots$ & 2.02$\pm$0.02  & 3.97$\pm$0.03  & 16725.62 \\
	&   &   &   &  HF-LL  & 1.25 & $\ldots$ & 16.4$\pm$0.6 & 0.03$\pm$0.002 & 3.24 \\
	20190711A 	&  2  & 593.1 & 0.632 & HF-HL  & 1.25 &  $\ldots$  & 6.5$\pm$0.5  & 34$\pm$3  & 1132.29 \\
	&   &    &   &  HF-LL  & 2.35 &  1.4$\pm$0.2 & 1$\pm$0.1  & 1.4$\pm$0.1 & 569.74 \\
	20200120E	&  80  & 87.82 & 0.022$^{a}$ & HF-HL  & 2.25 & 59$\pm$12 & 0.033$\pm$0.001  & 0.76$\pm$0.15  & 14.76 \\
	&  &  &   &  HF-LL  & 1.4 & $\ldots$ & 0.405$\pm$0.001 & 0.04$\pm$0.01 & 0.02 \\
	&   &  &  & LF-HL  & 0.6 & 1.7$\pm$1.1 & 0.07$\pm$0.1  & 2.4$\pm$1.4 & 0.11 \\
	&  &   &  &  LF-LL  & 0.6 & 0.06$\pm$0.06 & 0.41$\pm$0.04 & 0.50$\pm$0.03 & 0.0041 \\
	20201124A  & 2883  & 410.83 & 0.16 & HF-HL  & 1.25 & 52$\pm$6 & 12.1$\pm$0.2  & 640$\pm$70  & 462.43 \\
	&  &   &  &  HF-LL  & 1.25 & 0.009$\pm$0.002 & 5.30$\pm$1.56 & 0.05$\pm$0.01 & 0.076 \\
	&   &   &  & LF-HL  & 0.6 & \textgreater3$\pm$0.7 & 5.48$\pm$0.11  & \textgreater140$\pm$47 & 12.81 \\
	&  &   &    &  LF-LL  & 0.6 & 0.27$\pm$0.17 & 6$\pm$2 & 3.2$\pm$0.8 & 1.15 \\
	20220912A & 1773  & 219.46 & 0.13 & HF-HL  & 1.25 & $\ldots$ & 3.08  & 37.74$\pm$0.46  & 65.81 \\
	&   &   &  &  HF-LL  & 1.25 & $\ldots$ & 0.58 & 0.01 & 0.09 \\
	20180814A & 22  & 189.4 & 0.06$^{a}$ & LF-HL & 0.6 & 3.2$\pm$3.1 & 7.6$\pm$1.6  & 46$\pm$32  & 1.51 \\
	&   &  &  &  LF-LL  & 0.6 & 0.33$\pm$0.2 & 3.78$\pm$0.38 & 1.72$\pm$0.7 & 0.16 \\
	20180908B & 4  & 195.6 & 0.09 & LF-HL & 0.6 & 1.17$\pm$0.5 & 1.88$\pm$0.09  & 5$\pm$1.7  & 1.48 \\
	&   &  &   &  LF-LL  & 0.6 & 0.2$\pm$0.12 & 8.6$\pm$1.4 & 0.88$\pm$0.26 & 0.25 \\
	20180909A & 2  & 408.65 & 0.37 & LF-HL & 0.6 & 0.43$\pm$0.24 & 6.31$\pm$0.93  & 0.9$\pm$0.5  & 12.43 \\
	&   &  &   &  LF-LL  & 0.6 & 0.34$\pm$0.21 & 0.35$\pm$0.25 & 0.8$\pm$0.39 & 9.83 \\
	20180910A & 3  & 684.41 & 0.74  & LF-HL & 0.6 & 6.5$\pm$3.2 & 0.21$\pm$0.03  & 5.6$\pm$3 & 994.62 \\
	&   &  &  &  LF-LL  & 0.6 & 2.1$\pm$1 & 1.56$\pm$0.06 & 9.1$\pm$4.1 & 321.34 \\
	20181017A & 6  & 1281.6 & 1.605 & LF-HL & 0.6 & 0.4$\pm$0.2 & 20.2$\pm$1.7  & 16$\pm$5  & 417.55 \\
	&   & &   &  LF-LL  & 0.6 & 0.4$\pm$0.3 & 13.4$\pm$1.4 & 1$\pm$0.5 & 417.55 \\
	20181030A & 9  & 103.5 & 0.04$^{a}$ & LF-HL & 0.6 & 4.3$\pm$3.6 & 0.70$\pm$0.09  & 8.2$\pm$5.9  & 1.19 \\
	&   & &   &  LF-LL  & 0.6 & 0.45$\pm$0.35 & 1.33$\pm$0.19 & 1.79$\pm$0.7 & 0.12 \\
	20181119A & 17  & 364.05 & 0.33  & LF-HL & 0.6 & 1.11$\pm$0.35 & 2.54$\pm$0.09  & 5.4$\pm$1.5 & 25.09 \\
	&   & &    &  LF-LL  & 0.6 & 0.3$\pm$0.16 & 7.1$\pm$1.5 & 4.01$\pm$0.96 & 6.78 \\
	20181128A & 12  & 450.5  & 0.28 & LF-HL & 0.6 & 0.64$\pm$0.37 & 1.82$\pm$0.36  & 8.1$\pm$3.8 & 9.62 \\
	&   &  &   &  LF-LL  & 0.6 & 0.36$\pm$0.29 & 2.12$\pm$0.61 & 2.04$\pm$0.87 & 5.41 \\
	20181201D & 4  & 448.27 & 0.42 & LF-HL & 0.6 & 2.9$\pm$1.3 & 0.99$\pm$0.04  & 7.5$\pm$3.2  & 109.29 \\
	&   &  &    &  LF-LL  & 0.6 & 0.56$\pm$0.47 & 0.22$\pm$0.18 & 1.4$\pm$1.1 & 21.11 \\
	20181226F & 3  & 241.89 & 0.12 & LF-HL & 0.6 & 3$\pm$1.7 & 0.44$\pm$0.08  & 11.3$\pm$6.1  & 6.84 \\
	&   & &    &  LF-LL  & 0.6 & 0.99$\pm$0.63 & 0.51$\pm$0.05 & 1.8$\pm$1 & 2.26 \\
	20190107B & 2  & 166.09 & 0.07$^{a}$ & LF-HL & 0.6 & 2.8$\pm$1.7 & 0.45$\pm$0.02  & 4.3$\pm$2.5 & 1.79 \\
	&   &   &   &  LF-LL  & 0.6 & 1.11$\pm$0.87 & 0.52$\pm$0.32 & 1.39$\pm$0.85 & 0.71 \\
	20190110C & 3  & 221.96  & 0.13 & LF-HL & 0.6 & 0.64$\pm$0.39 & 0.75$\pm$0.04  & 1.4$\pm$0.76 & 1.74 \\
	&   &  &   &  LF-LL  & 0.6 & 0.23$\pm$0.10 & 0.9$\pm$0.3 & 1.25$\pm$0.27 & 0.62 \\
	20190113A & 3  & 428.92 & 0.11 & LF-HL & 0.6 & 1.8$\pm$1.2 & 3.03$\pm$0.45  & 9.4$\pm$4.5 & 3.21 \\
	&   &   &  &  LF-LL  & 0.6 & 1.1$\pm$0.97 & 1.82$\pm$0.21 & 5.7$\pm$4 & 1.96 \\
	20190116A & 2  & 428.92 & 0.45 & LF-HL & 0.6 & 0.4$\pm$0.2 & 1.5$\pm$0.3  & 2.8$\pm$1.4  & 18.38 \\
	&   &   &  &  LF-LL  & 0.6 & 0.3$\pm$0.2 & 4$\pm$0.5 & 0.8$\pm$0.4 & 13.78 \\
	20190117A & 21  & 393.6 & 0.36 & LF-HL & 0.6 & 3.74$\pm$0.85 & 2.24$\pm$0.14  & 43.6$\pm$9.4  & 98.14 \\
	&   &   &   &  LF-LL  & 0.6 & 0.6$\pm$0.2 & \textless1.3$\pm$0.29 & 6.7$\pm$1.3 & 15.75 \\
	20190127B & 2  & 663.03 & 0.73 & LF-HL & 0.6 & 3.9$\pm$1.7 & 1.176$\pm$0.00058  & 9.9$\pm$3.5 & 575.15 \\
	&   &   &  &  LF-LL  & 0.6 & 0.63$\pm$0.54 & 2.5$\pm$1.1 & 11.4$\pm$5.8 & 92.91 \\
	20190208A & 15  & 580.05  & 0.58 & LF-HL & 0.6 & 0.58$\pm$0.19 & 1.21$\pm$0.11  & 1.62$\pm$0.34 & 49.09 \\
	&   &   &  &  LF-LL  & 0.6 & 0.39$\pm$0.27 & 6.06$\pm$0.63 & 4.9$\pm$2.9 & 33.01 \\
	20190209A & 2  & 425 & 0.40 &  LF-HL & 0.6 & 0.77$\pm$0.55 & 1.43$\pm$0.26  & 3.9$\pm$2.3 & 26.75 \\
	&   &   &   &  LF-LL  & 0.6 & 0.25$\pm$0.21 & 6.73$\pm$0.43 & 1.81$\pm$0.2 & 8.69 \\
	20190210C & 2  & 643.37 & 0.69 & LF-HL & 0.6 & 2.37$\pm$0.81 & 0.29$\pm$0.05  & 3.6$\pm$1.2 & 302.09 \\
	&   &   &  &  LF-LL  & 0.6 & 0.8$\pm$0.51 & 1.39$\pm$0.23 & 2.4$\pm$1.3 & 101.97 \\
	20190212A & 13  & 302 & 0.22 & LF-HL & 0.6 & 0.92$\pm$0.5 & 0.28$\pm$0.16  & 4.2$\pm$1.8  & 7.58 \\
	&   &   &  &  LF-LL  & 0.6 & 0.33$\pm$0.25 & 0.66$\pm$0.14 & 0.83$\pm$0.31 & 2.72 \\
	20190213A & 2  & 651.45 & 0.72 & LF-HL & 0.6 & 0.55$\pm$0.36 & 1.77$\pm$0.74  & 4.1$\pm$1.8  & 79.48 \\
	&   &   &  &  LF-LL  & 0.6 & 0.45$\pm$0.34 & 1.51$\pm$0.5 & 2.3$\pm$1.2 & 65.03 \\
	20190222A & 2  & 460.6 & 0.36 & LF-HL & 0.6 & 1.83$\pm$0.73 & 1.24$\pm$0.54  & 8.6$\pm$3.3  & 50.60 \\
	&   &   &  &  LF-LL  & 0.6 & 1.70$\pm$0.87 & 1.84$\pm$0.07 & 11.8$\pm$5.5 & 47.00 \\
	20190226B & 3  & 631.60 & 0.68 & LF-HL & 0.6 & 0.81$\pm$0.55 & 0.97$\pm$0.11  & 2.8$\pm$1.9  & 101.41 \\
	&   &   &  &  LF-LL  & 0.6 & 0.2$\pm$0.17 & 5.9$\pm$1 & 2.1$\pm$1.1 & 25.04 \\
	20190303A & 38  & 222.4  & 0.14 & LF-HL & 0.6 & 2.96$\pm$0.94 & 0.89$\pm$0.03  & 10.4$\pm$3.1 & 9.58 \\
	&   &   &  &  LF-LL  & 0.6 & 0.2$\pm$0.12 & 7.2$\pm$1.3 & 2.2$\pm$1 & 0.65 \\
	20190303D & 2  & 711.15  & 0.816 & LF-HL & 0.6 & 1.8$\pm$1.2 & 0.76$\pm$0.1  & 1.13$\pm$0.69 & 350.57 \\
	&   &  &   &  LF-LL  & 0.6 & 0.59$\pm$0.35 & 0.81$\pm$0.1 & 1.17$\pm$0.67 & 114.91 \\
	20190328C & 2  & 472.86 & 0.43 & LF-HL & 0.6 & 4.7$\pm$2.3 & 0.77$\pm$0.14  & 14.9$\pm$7.1  & 190.61 \\
	&   &  &   &  LF-LL  & 0.6 & 0.92$\pm$0.55 & 1.53$\pm$0.86 & 12.9$\pm$5.3 & 37.31 \\
	20190417A & 19  & 1378.2  & 1.67 & LF-HL & 0.6 & 0.76$\pm$0.25 & 0.80$\pm$0.12  & 3.48$\pm$0.84 & 880.42 \\
	&   &   &  &  LF-LL  & 0.6 & 0.32$\pm$0.12 & 1.25$\pm$0.29 & 1.84$\pm$0.37 & 370.70 \\
	20190430C & 3  & 400.56 & 0.31 & LF-HL & 0.6 & 5.8$\pm$2.0 & 0.89$\pm$0.095  & 15.2$\pm$5.2  & 106.15 \\
	&   &  &  &  LF-LL  & 0.6 & 1.9$\pm$1.2 & 3.38$\pm$0.62 & 10.1$\pm$3.9 & 34.78 \\
	20190604A & 3  & 552.65 & 0.60 & LF-HL & 0.6 & 0.92$\pm$0.41 & 1.84$\pm$0.12  & 7.8$\pm$2.5  & 84.55 \\
	&   &  &  &  LF-LL  & 0.6 & 0.48$\pm$0.27 & 1.33$\pm$0.2 & 1.87$\pm$0.9 & 44.11 \\
	20190609C & 3  & 480.28 & 0.32 & LF-HL & 0.6 & 3.3$\pm$1.5 & 0.51$\pm$0.046  & 4.7$\pm$1.4  & 67.51 \\
	&   &  &  &  LF-LL  & 0.6 & 0.64$\pm$0.21 & 2.07$\pm$0.25 & 1.91$\pm$0.43 & 13.09 \\
	20190804E & 8  & 363.68  & 0.32 & LF-HL & 0.6 & 1.2$\pm$0.32 & 0.99$\pm$0.02  & 4.83$\pm$0.75 & 24.37 \\
	&   &  &    &  LF-LL  & 0.6 & 0.26$\pm$0.1 & 4.62$\pm$0.35 & 1.35$\pm$0.2 & 5.28 \\
	20190812A & 2  & 254.52 & 0.16 & LF-HL & 0.6 & 4.3$\pm$1.5 & 0.60$\pm$0.02  & 13$\pm$4.4  & 18.10 \\
	&   &  &  &  LF-LL  & 0.6 & 1.17$\pm$0.65 & 0.42$\pm$0.06 & 0.65$\pm$0.37 & 4.93 \\
	20190905A & 6  & 254.11 & 0.14 & LF-HL & 0.6 & 7.8$\pm$3.5 & 0.7451$\pm$0.0051  & 18$\pm$7.9  & 24.47 \\
	&   &   &   &  LF-LL  & 0.6 & 0.37$\pm$0.28 & 1.13$\pm$0.16 & 1.79$\pm$0.68 & 1.16 \\
	20190907A & 7 & 309.6 & 0.22 & LF-HL & 0.6 & 0.4$\pm$0.2 & 0.54$\pm$0.14  & 0.9$\pm$0.4  & 3.61 \\
	&   &  &  &  LF-LL  & 0.6 & 0.2$\pm$0.1 & 3$\pm$0.8 & 0.7$\pm$0.2 & 1.81 \\
	20190915D & 7 & 488.69 & 0.42 & LF-HL & 0.6 & 2.56$\pm$0.98 & 4.99$\pm$0.43  & 32$\pm$10 & 99.27 \\
	&   &  &  &  LF-LL  & 0.6 & 0.3$\pm$0.19 & 9.7$\pm$1.2 & 2.67$\pm$0.94 & 11.63 \\
	20191013D & 2 & 523.57 & 0.54 & LF-HL & 0.6 & 4.2$\pm$1.6 & 2.44$\pm$0.36  & 50$\pm$16  & 302.99 \\
	&   &  &  &  LF-LL  & 0.6 & 1.08$\pm$0.71 & 4.79$\pm$0.52 & 7.2$\pm$4 & 77.91 \\
	20191105B & 2 & 318.61 & 0.18 & LF-HL & 0.6 & 7.6$\pm$3.9 & 0.55$\pm$0.02  & 19.7$\pm$9.8 & 41.57 \\
	&   &  &  &  LF-LL  & 0.6 & 1.89$\pm$0.95 & 0.71$\pm$0.05 & 2.7$\pm$1.3 & 10.34 \\
	20191106C & 20 & 333.4 & 0.30 & LF-HL & 0.6 & 0.18$\pm$0.06 & 5.12$\pm$0.45 & 1.48$\pm$0.26  & 3.07 \\
	&   &   &  &  LF-LL  & 0.6 & 0.071$\pm$0.045 & 3.90$\pm$0.65 & 0.94$\pm$0.19 & 1.23 \\
	20191114A & 3 & 552.47  & 0.52 & LF-HL & 0.6 & 1.2$\pm$1 & 4.35$\pm$0.45 & 3.4$\pm$1.5 & 78.71 \\
	&   &  &  &  LF-LL  & 0.6 & 0.43$\pm$0.31 & 4.68$\pm$0.27 & 3.8$\pm$1.9 & 28.20 \\
	20200118D & 2 & 625.57 & 0.61 & LF-HL & 0.6 & 1.05$\pm$0.27 & 1.28$\pm$0.12 & 2.98$\pm$0.49 & 99.26 \\
	&   &  &   &  LF-LL  & 0.6 & 0.12$\pm$0.06 & 0.85$\pm$0.18 & 0.8$\pm$0.2 & 11.16 \\
	20200127B & 2 & 351.3 & 0.28 & LF-HL & 0.6 & 5.6$\pm$2.1 & 0.61$\pm$0.03 & 7.7$\pm$2.8 & 84.13 \\
	&   &  &  &  LF-LL  & 0.6 & 2.41$\pm$0.89 & 0.45$\pm$0.02 & 4.7$\pm$1.6 & 36.21 \\
	20200202A & 4 & 722.37 & 0.76 & LF-HL & 0.6 & 3$\pm$1.4 & 1.49$\pm$0.02 & 32$\pm$14  & 482.31 \\
	&   &   &   &  LF-LL  & 0.6 & 0.28$\pm$0.19 & 14.87$\pm$0.84 & 5.7$\pm$2.8 & 45.02 \\
	20200223B & 10 & 202.27 & 0.09 & LF-HL & 0.6 & 7.5$\pm$5.4 & 2.24$\pm$0.14 & 14.4$\pm$2.7 & 9.68 \\
	&   &  &  &  LF-LL  & 0.6 & 0.54$\pm$0.23 & 1.34$\pm$0.12 & 1.06$\pm$0.36 & 0.70 \\
	20200320A & 2 & 594.93 & 0.64 & LF-HL & 0.6 & 19$\pm$6.6 & 1.19$\pm$0.02 & 114$\pm$37 & 2028 \\
	&   &  &  &  LF-LL  & 0.6 & 0.63$\pm$0.39 & 0.60$\pm$0.09 & 1.26$\pm$0.85 & 67.24 \\
	20200420A & 2 & 671.5 & 0.731 & LF-HL & 0.6 & 6.1$\pm$2.9 & 0.84$\pm$0.02 & 22$\pm$10  & 905.68 \\
	&   &   &  &  LF-LL  & 0.6 & 0.58$\pm$0.2 & 0.944$\pm$0.065 & 2.02$\pm$0.44 & 86.11 \\
	20200508H & 2 & 479.49 & 0.45 & LF-HL & 0.6 & 0.91$\pm$0.63 & 0.66$\pm$0.05 & 2.4$\pm$2  & 42.26 \\
	&   &  &  &  LF-LL  & 0.6 & 0.39$\pm$0.27 & 3.93$\pm$0.5 & 0.93$\pm$0.42 & 18.11 \\
	20200619A & 6 & 439.77 & 0.41 & LF-HL & 0.6 & 0.76$\pm$0.22 & 0.49$\pm$0.08 & 1.76$\pm$0.48  & 28.48 \\
	&   &   &  &  LF-LL  & 0.6 & 0.175$\pm$0.097 & 0.002$\pm$0.075 & 1.05$\pm$0.22 & 6.56 \\
	20200809E & 4 & 1703.48 & 2.13 & LF-HL & 0.6 & 1.22$\pm$0.77 & 2.93$\pm$0.14 & 7$\pm$3.8  & 2543.27 \\
	&   &  &  &  LF-LL  & 0.6 & 0.41$\pm$0.22 & 4.31$\pm$0.88 & 1.48$\pm$0.26 & 854.71 \\
	20200828A & 2 & 561.31 & 0.57 & LF-HL & 0.6 & 0.59$\pm$0.33 & 3.11$\pm$0.57 & 3.4$\pm$1.3  & 48.48 \\
	&   &   &  &  LF-LL  & 0.6 & 0.49$\pm$0.29 & 2.31$\pm$0.39 & 10.9$\pm$4 & 40.27 \\
	20200913C & 2 & 576.88  & 0.61 & LF-HL & 0.6 & 0.7$\pm$0.37 & 0.54$\pm$0.03 & 0.54$\pm$0.25 & 66.44 \\
	&   &   & &  LF-LL  & 0.6 & 0.68$\pm$0.23 & 0.731$\pm$0.083 & 1.6$\pm$0.36 & 64.54 \\
	20200926A & 6 & 758.45 & 0.84 & LF-HL & 0.6 & 0.34$\pm$0.24 & 2.38$\pm$0.26 & 0.94$\pm$0.46  & 71.14 \\
	&   &   &  &  LF-LL  & 0.6 & 0.16$\pm$0.12 & 2.11$\pm$0.27 & 0.56$\pm$0.23 & 33.48 \\
	20200929C & 16 & 413.66 & 0.40 & LF-HL & 0.6 & 0.91$\pm$0.47 & 0.62$\pm$0.09 & 4.8$\pm$1.6  & 31.24 \\
	&   &   &  &  LF-LL  & 0.6 & 0.46$\pm$0.27 & 1.7$\pm$0.13 & 2.75$\pm$0.89 & 15.79 \\
	20201114A & 2 & 322.23 & 0.268 & LF-HL & 0.6 & 0.72$\pm$0.51 & 1.04$\pm$0.31 & 4.7$\pm$2.2 & 9.78 \\
	&   &   &  &  LF-LL  & 0.6 & 0.45$\pm$0.27 & 0.72$\pm$0.21 & 2.45$\pm$0.73 & 6.11 \\
	20201130A & 12 & 287.98 & 0.166 & LF-HL & 0.6 & 6.8$\pm$6.3 & 3.18$\pm$0.45 & 18.4$\pm$7.7 & 31.49 \\
	&   &  &  &  LF-LL  & 0.6 & 0.45$\pm$0.21 & 3.32$\pm$0.55 & 1.31$\pm$0.68 & 2.08 \\
	20201221B & 6 & 510.42  & 0.511 & LF-HL & 0.6 & 0.58$\pm$0.18 & 2.62$\pm$0.22 & 6.4$\pm$1.6 & 35.97 \\
	&   & &   &  LF-LL  & 0.6 & 0.19$\pm$0.13 & 5$\pm$0.49 & 1.19$\pm$0.32 & 11.78 \\
	20210323C & 11 & 288.48 & 0.201 & LF-HL & 0.6 & 1.6$\pm$1.1 & 5$\pm$0.25 & 7.6$\pm$3.1  & 11.33 \\
	&   &  &  &  LF-LL  & 0.6 & 0.8$\pm$0.41 & 0.83$\pm$0.07 & 2.32$\pm$0.71 & 5.67 \\
\end{longtable}

\tablefoot { The redshift $z$ in the 4th column are estimated by referring to the YMW16 model \citep{Yao+17}, where we fix DM$_{host}$ = 100 pc cm$^{-3}$. For FRBs' redshift $z$ marked with a letter $a$, the DM$_{host}$ is assumed to be 40 pc cm$^{-3}$ in order to ensure that the estimated redshift $z$ is larger than zero. [1] The 2nd column of Table \ref{table1} refers to the number of observed bursts for each repeating FRB. [2] The DM in the 3rd column of the Table \ref{table1} refers to the observe dispersion measure obtained by maximizing S/N.  }
\clearpage
%%%%%%%%%%%%%%%table2%%%%%%%%%%%%%%%%%
\begin{table}
	\centering
	\caption{ The instrumental parameters of telescopes }
	{\renewcommand{\arraystretch}{1.5}
		\begin{tabular}{ccccccc}
			\hline\hline
			{Telescope} & {$\nu_c$} & {Bandpass} & {$F_{\rm th}$} & { $\Omega$ } & {$T$} & {Ref.}  \\
			{} & {(GHz)} & {(GHz)} & {(Jy)} & {(deg$^{2}$)} & {(Years)} & {}   \\
			\hline
			GBT& 1.5&1.0& 0.13 Jy & 0.07 & 23& [1] \\
			&6.0&4.0& 0.224 Jy &0.08 & 23 &  [1] \\
			&3.0&2.0& 4.454 Jy & 0.10 & 23 &  [1]  \\
			&22.5&9.0 &1.856 Jy& 0.09 & 23 &   [1] \\
			\hline
			VLA&  1.43& 1.0&50 $\mu$Jy & 0.88 & 43 & [2]  \\
			&	 4.86 &4.0& 20 $\mu$Jy & 0.08  & 43 &  [2]  \\
			&	 8.46& 4.0 &13 $\mu$Jy & 0.04 & 43 &  [2]  \\
			\hline
			FAST & 1.45&0.57& 2 $\mu$Jy & 0.01  & 5 & [2] \\
			\hline
			ASKAP& 1.4&0.3& 60 $\mu$Jy & 120 & 11 & [2] \\
			\hline
			CHIME &0.6& 0.4&160-270 mJy & 134  & 5 & [3]  \\
			\hline
			Parkes &1.5&2.0 & 180-250 mJy & 0.56 & 62 & [4]  \\
			\hline
		\end{tabular}
	}
	\tablefoot{ The parameters are taken from the following references: [1] \cite{Surnis+19}. [2] \cite{ZhangZB+18}. [3] \cite{Chime+23}. [4] \cite{Maan+17}.  }
	\label{table2}
\end{table}

%%%%%%%%%%%%%%%%%%%%%%%table3%%%%%%%%%%%%%%%%%%%

\begin{table}
	\centering
	\caption{The best fitted parameters of the luminosity functions}
	\resizebox{\columnwidth}{!}{
		\begin{tabular}{ccccc}
			\hline\hline
			Parameters & HF-HL & HF-LL & LF-HL & LF-LL \\
			\hline
			Model & SPL & SPL & BPL & BPL \\
			\hline
			$L_{b}$ (erg/s) & $\cdots$ & $\cdots$ & (2.09$\pm$0.18)$\times$10$^{42}$ & \makebox[0.25\textwidth]{ (1.73$\pm$0.31)$\times$10$^{42}$ } \\
			$\alpha_{1}$ & -1.04$\pm$0.02 & -1.07$\pm$0.02 & 0.49$\pm$0.05  & 0.49$\pm$0.07 \\
			$\alpha_{2}$ & $\cdots$ & $\cdots$ & 1.36$\pm$0.05  & 1.67$\pm$0.19 \\
			$\omega$ & $\cdots$ & $\cdots$ & 3.50 & 1.17 \\
			$\chi^2_{\nu}$&1.68&1.00 &1.79	&2.49\\
			\hline
	\end{tabular} }
	\label{table3}
\end{table}
%%%%%%%%%%%%%%%%table4%%%%%%%%%%%%%%%%%%%
\begin{table}
	\centering
	\caption{ The 2D K-S test results of local event rates among diverse samples }
	{\renewcommand{\arraystretch}{1.5}
		\begin{tabular}{c|cccccccc}
			\hline\hline
			Event rate	& Pairs &  $D$ &  $D_{\alpha}$ &  $p$-value &  $N_{1}$ & $N_{2}$ & $\alpha$ & Same \\
			\hline
			& HFHL-HFLL & 0.63 & 0.36 & 2.18$\times$10$^{-5}$ & 25 & 35 & 0.05 & N \\
			& HFLL-LFHL & 0.28 & 0.29 & 6.42$\times$10$^{-2}$ & 55 & 35 & 0.05 & Y \\
			log$\rho_{0,L}$ & HFHL-LFLL & 0.74 & 0.35& 1.02$\times$10$^{-7}$ & 25 & 40 & 0.05 & N \\
			& LFHL-LFLL & 0.26 & 0.28 & 9.17$\times$10$^{-2}$ & 55 & 40 & 0.05 & Y \\
			& HFHL-LFHL & 0.66 & 0.33 & 7.10$\times$10$^{-7}$ & 55 & 25 & 0.05 & N \\
			& HFLL-LFLL & 0.18 & 0.31 & 0.61 & 35 & 40 & 0.05 & Y \\
			\hline
			& HFHL-HFLL & 0.68 & 0.36 & 7.63$\times$10$^{-6}$ & 25 & 35 & 0.05 & N \\
			& HFLL-LFHL & 0.26 & 0.29 & 0.11 & 55 & 35 & 0.05 & Y \\ 
			log$\rho_{0,>L}$ & HFHL-LFLL & 0.77 & 0.35 & 2.69$\times$10$^{-8}$ & 25 & 40 & 0.05 & N \\ 
			& LFHL-LFLL & 0.25 & 0.28 & 0.11 & 55 & 40 & 0.05 & Y \\
			& HFHL-LFHL & 0.68 & 0.33 & 2.94$\times$10$^{-2}$ & 55 & 25 & 0.05 & N \\
			& HFLL-LFLL & 0.17 & 0.31 & 0.64 & 35 & 40 & 0.05 & Y \\ 
			\hline
		\end{tabular}
	}
	\tablefoot{ Y and N indicate that the event rates of two samples are consistent and inconsistent, correspondingly. }
	\label{table4}
\end{table}

%%%%%%%%%%%%%%%%%%%%%%table5%%%%%%%%%%%%%%%%%%%%%%
\begin{table}
	\centering
	\caption{ The best fitted parameters of the rates of the four sub-samples of FRBs }
	{\renewcommand{\arraystretch}{1.5}
		\begin{tabular}{ccccccc}
			\hline\hline
			{Parameters}  &  {HF-HL} & {HF-LL} & {LF-HL} & {LF-LL} & {LF$_{BS}$} & {HF$_{BS}$} \\
			\hline
			A & 4.53$\pm$3.9 & 1.63$\pm$0.8 & 19.98$\pm$5.0 & 17.09$\pm$4.8 & 33.34$\pm$8.7 & 3.94$\pm$0.4 \\
			\hline
			B & 1.25$\pm$0.1 & 1.27$\pm$0.02 & 1.16$\pm$0.02 & 1.20$\pm$0.02 & 1.17$\pm$0.02 & 1.25$\pm$0.01 \\
			\hline
			C & -15.79$\pm$2.0 & -27.55$\pm$2.4 & -36.38$\pm$6.3 & -28.44$\pm$3.0 & -29.22$\pm$4.5 & -20.59$\pm$0.5 \\
			\hline
			D & -10.81$\pm$1.4 & -8.98$\pm$1.8 & -7.57$\pm$0.5 & -6.82$\pm$0.5 & -6.55$\pm$0.5 & -9.08$\pm$0.3 \\ 
			\hline
		\end{tabular}
	}
	\tablefoot{ Symbols LF$_{BS}$ and HF$_{BS}$ represent the low- and high-frequency data resampled with the bootstrap method. }
	\label{table5}
\end{table}
%%%%%%%%%%%%%%%%%%%%%%%%%%%%%%%%%%%%%%%%%%%%%%%%%%%%%%%%%%%%%%%%%%%%%%%%%%%%%%%%%%
\end{document}